\documentclass[aps,preprint,preprintnumbers, amsmath, amssymb, prb,
showpacs,superscriptaddress]{revtex4-1}
\usepackage{times}
\usepackage{graphicx}
\usepackage{psfrag}
\usepackage{ae}
\usepackage{amsmath,amssymb}
\usepackage[usenames]{color}
\usepackage{float}

\begin{document}

\author{Leandro B. Krott} 
\email{leandro.krott@ufrgs.br}
\affiliation{Instituto 
de F\'{\i}sica, Universidade Federal
do Rio Grande do Sul, Caixa Postal 15051, CEP 91501-970, 
Porto Alegre, RS, Brazil}

\author{Jos\'e Rafael Bordin} 
\email{josebordin@unipampa.edu.br}
\affiliation{Campus Ca\c capava do Sul, Universidade Federal
do Pampa, Av. Pedro Anuncia\c c\~ao, s/n, CEP 96570-000, 
Ca\c capava do Sul, RS, Brazil}

\author{Ney Mar\c cal Barraz Jr} 
 \email{ney.barraz@uffs.edu.br}
\affiliation{Campus Cerro Largo, Universidade Federal
da Fronteira Sul, Av. Jacob Reinaldo Haupenthal, 1580. CEP 97900-000, 
Cerro Largo, RS, Brazil}

\author{Marcia C. Barbosa} 
\email{marciabarbosa@ufrgs.br}
\affiliation{Instituto 
de F\'{\i}sica, Universidade Federal
do Rio Grande do Sul, Caixa Postal 15051, CEP 91501-970, 
Porto Alegre, RS, Brazil}

\title{Effects of nanopore and fluid structure on anomalies 
and phase transitions of confined core-softened fluids}
\date{\today}
\begin{abstract}

We use Molecular Dynamics simulations to study
how the nanopore and
the fluid structures affects the dynamic, thermodynamic and structural 
properties of a 
confined anomalous fluid. The fluid is modeled using an effective pair
 potential derived from the ST4 atomistic model for water. This
system exhibits
 density, structural and dynamical anomalies and the vapor-liquid
and liquid-liquid critical points similar to the
quantities observed in bulk water.  The confinement
is modeled both by smooth and structured walls. The temperatures
of extremum  density and
diffusion  for the confined fluid
show a shift to lower values while
the pressures move to higher amounts for
both smooth and structured confinement. In the 
case of smooth walls, the critical points and the limit between 
fluid and amorphous phases show a non-monotonic change in
the temperatures and pressures when the 
the nanopore size is increase. In the
case of  structured walls the pressures and 
temperatures of the critical points varies monotonicaly with
the porous size. Our results are explained on
basis of the competition between the different length scales
of the fluid and the wall-fluid interaction.

\end{abstract}
\maketitle

\section{\label{sec1}Introduction}

Water is an important material in industry, technology and biological 
processes
due to its unusual properties. 
Water unusual properties comprise many anomalous behavior, with 70 
known anomalies~\cite{URL},
like the maximum value of its density in $T = 4^o$C
at room pressure, and the increase of the diffusion as the system is 
compressed~\cite{Wa64,Ke75,An76}.
These anomalies have been explained in terms of the formation
of hydrogen bond network. The water molecules form open and compact (bonded
and nonbonded) clusters of tetramers. From the competition
between these structures the anomalies arise.

As a natural consequence of the polimorphism of water clusters
the pressure-temperature phase diagram of water is very complex. At 
low temperatures,
water shows a coexistence of two amorphous phases:
a low density amorphous and a high density amorphous. For higher temperatures, 
these two amorphous phases might lead to the appearance of  two liquid phases,
separated by a first order phase transition line ended in a liquid-liquid
critical point ($LLCP$). Whereas, homogeneous nucleation occurs in 
this region, that is called {\it no man's land}, and because of that, it 
is a incredible hard task to do experimental 
measures of liquid water in bulk systems in this region. 
Theoretically, the existence of these two 
liquid phases was evidenced in the atomistic ST4 model by Poole 
and co-authors~\cite{Po92}
and confirmed in recent simulations~\cite{Liu12, Palmer14}. As well, new
experiments shows a evidence of the coexistence of a high-density and 
low-density liquid phase of water~\cite{Taschin13}. 
A $LLCP$ was also predicted for others atomistic models
of water~\cite{Brovchenko05,Poole11,Poole13,Kesselring13,Starr14,Yagasaki14}, 
and in models for 
phosphorus~\cite{Morishita01}, silica~\cite{Voivod01,Lascaris14}, 
silicon~\cite{Vasisht11}, carbon~\cite{Glosli99}, hydrogen~\cite{Scandolo03}
and colloidal systems~\cite{Smallenburg14}. 
On the other hand, recent studies suggests that the $LLCP$ 
can be a open trend on 
supercooled water and 
other materials.~\cite{Limmer11,Limmer13, Angell14}. In this way, 
there is still several open questions about the $LLCP$.

As an attempt to avoid the crystallization of 
water in the {\it no man's land}, experiments 
with nanoconfined water have been performed 
recently~\cite{Li05, Nagoe14, Mallamace14}. 
The presence of a confining structure changes the number of hydrogen 
bonds, avoiding the nucleation. 
Some experiments of water confined in nanopores, performed by x-ray 
and neutron scattering, show 
that liquid states persist down to temperatures much lower 
than in bulk~\cite{Ja08, De10,Er11}.
The nanopores size has important influence in the crystallization of the 
system~\cite{MoK99,HwC07,Ja08,KiS09,Er11}, and hydrophilic and hydrophobic
nanopores can lead to distinct results as 
well~\cite{De10,faraone_liu_chen_2009}.

Classical atomistic models for water
are important tools to understand its properties. 
On the other hand, coarse-grained models arise as an interesting tool 
to see the universal mechanisms that lead to anomalous waterlike properties.
Coarse-grained models may reproduce diffusion and density anomalies and can 
be modeled by
core-softened (CS) potentials with two length scales, that can be constructed 
using a shoulder or a 
ramp potential~\cite{He70,Ja98,Xu05,Ol06a,Lo07,
jonatas_evy_alan_marcia_2010}. 
These coarse-grained models for anomalous fluids are able to capture the bulk 
 waterlike anomalies and averaged properties
in the confined materials. When confined by fixed 
hydrophilic plates, the fluid-wall interaction can induce 
solidification and shift the anomalous properties to higher temperatures, 
while hydrophobic nanopores lead the system to remain in 
liquid state and shift the waterlike anomalies to
lower temperatures in relation to bulk~\cite{Kr13a,Kr14a}. Whereas, when the
 nanopore has
at least one degree of freedom, the anomalous behavior of the fluid disappear
 and distinct phase transitions
are observed~\cite{Kr13b,BoKr14a, Bordin14c, BoK15a}. CS fluids confined in 
nanotubes also present 
interesting findings, similar to obtained in atomistic models for water, as 
the increase in diffusion coefficient
and flux for narrow nanotubes associated to a layer to single-file transition 
and a discontinuity in the enhancement flow 
factor~\cite{Bordin12b,Bordin13a,Bordin14a}. The drawback of 
these core-softened potentials
is that due to the simplicity of the two length  scales, they
are not capable to reproduce the effects related to the
third coordination shell of the anomalous fluid what might be
relevant under confinement~\cite{Ho07}. 

In addition to the relevance
of the detail structure of the 
liquid, the structure
of the confining system is 
also relevant since biological and physical materials do not
exhibit the smoothness and regularity of the flat walls and
tubes enployed in the simulations.
This naturaly raises the question 
of what is the  role played by the structure of the 
liquid and of the 
interface in the thermodynamic, dynamic
and structural behavior of confined systems.
Recent simulations have shown that the hydrophobic or hydrophilic 
behavior of the confining surfaces 
are governed by the interfacial free energy, that strongly depends on
 the surface structure~\cite{Mi10}. 
Even thought these simulations do not observe
 important differences in the diffusion of the systems 
confined between smooth and rough walls~\cite{WK11}, they 
show that the adsorption 
behavior and the solvation pressure are significantly affected by the 
roughness of the  
confining surface~\cite{KYa11} and that different
liquid and solid phases that exist in the smooth confined
are not present in the rough case~\cite{BoKr14a}.

In this work we address the question of which are the 
effects of the roughness of the nanopore wall in the physical 
properties of a anomalous fluids. Our analysis is 
done in the framework of 
an effective model that incorporates not only
the two length scales traditionally present
in the CS potentials but additional
length scales representing the third coordination shell 
of the fluid. Our goal is also to understand the effect of
the structure of the liquid in the thermodynamic, 
dynamic and structural properties of a fluid confined in a
nanopore.

The paper is organized as follows: in Sec. II we introduce the model and 
the methods and simulation details are described; the 
results are given in Sec. III; and  conclusions are presented in Sec. IV.

\section{\label{sec:model} The model and simulation details}

In this paper all physical quantities are computed
in the standard LJ units~\cite{AllenTild},
\begin{equation}
\label{red1}
r^*\equiv \frac{r}{r_0}\;,\quad \rho^{*}\equiv \rho r_0^{3}\;, \quad 
\mbox{and}\quad t^* \equiv t\left(\frac{\gamma}{mr_0^2}\right)^{1/2}\;,
\end{equation}
for distance, density of particles and time , respectively, and
\begin{equation}
\label{rad2}
p^*\equiv \frac{p r_0^{3}}{\gamma}\;, \quad U^* \equiv \frac{U}{\gamma} \quad \mbox{and}\quad 
T^{*}\equiv \frac{k_{B}T}{\gamma}
\end{equation}
for the pressure, energy and temperature, respectively, where $r_0$ is the distance
parameter, $\gamma$ the energy parameter and $m$ the mass parameter.
Since all physical quantities are defined in reduced LJ units, 
the $^*$ is  omitted, in order to simplify the discussion.

\begin{figure}[!htb]
  \begin{centering}
\includegraphics[clip=true,width=8cm]{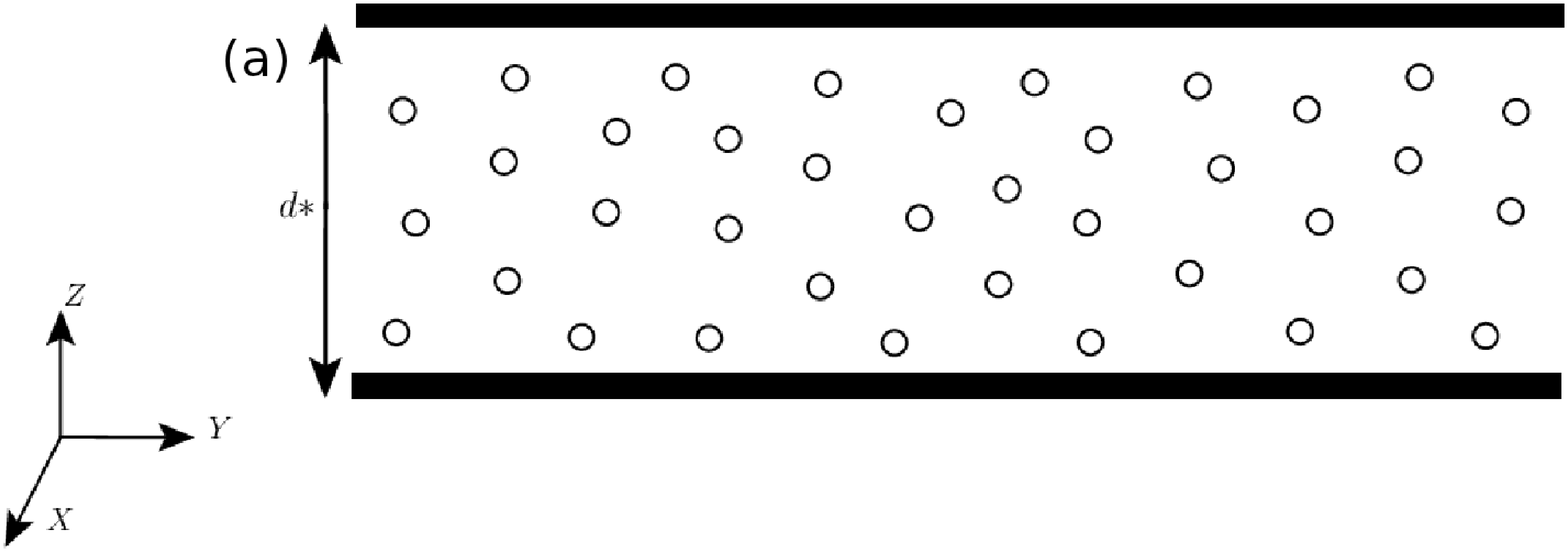}
\includegraphics[clip=true,width=8cm]{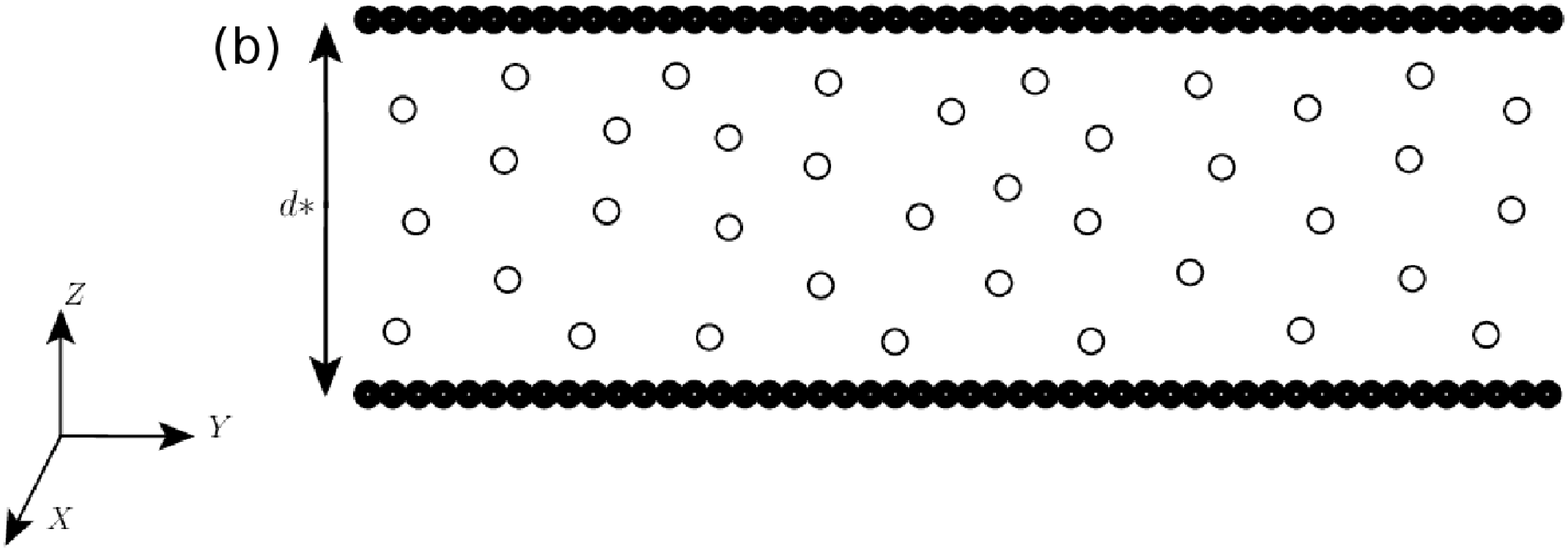}
  \end{centering}
  \caption{Schematic depiction of the particles confined between (a) 
smooth and 
(b) structured plates.}
\label{fig1}
\end{figure}

The fluid is composed by $N$ spherical particles 
of diameter $\sigma = 1.47$ and mass $m$ confined between two 
parallel and fixed plates.
We have studied two kinds of nanopores: with smooth and structured walls.
Smooth plates are modeled by force fields and do not 
have structure, interacting continuously 
with the fluid. Structured plates are formed by spherical 
particles in a square lattice 
with punctual interactions. A schematic depiction for the systems 
with (a) smooth 
and (b) structured plates is shown in Fig.~\ref{fig1}.

\begin{figure}[!htb]
  \begin{centering}
\includegraphics[clip=true,width=9cm]{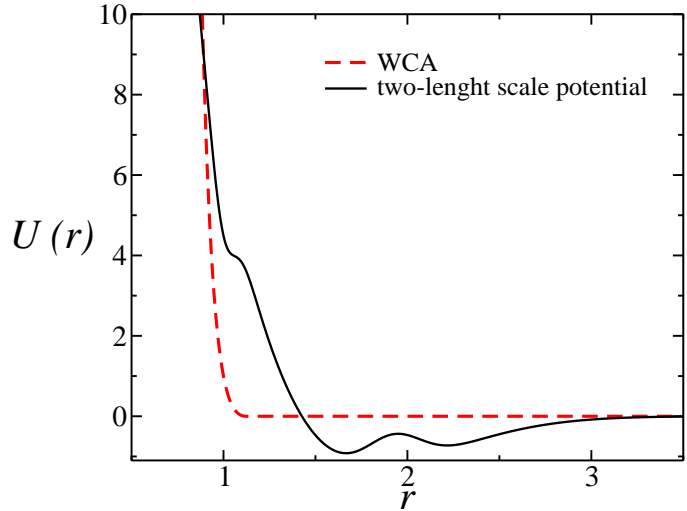}\par
    \par
  \end{centering}
  \caption{Particle-particle interaction potential (black solid line) 
and particle-plate interaction potential (red dashed line).}
  \label{fig2}
\end{figure}

The particles of the fluid interact through the isotropic effective 
potential~\cite{ney_evy_marcia_2009} given by 
\begin{equation}\label{eq_potential}
 \centering
 \frac{U(r)}{\epsilon} =
  \left[ \left( \frac{\sigma}{r} \right)^{a} 
- \left( \frac{\sigma}{r} \right)^{b} \right] +
  \sum_{j=0}^{4}h_j\exp\left[-\left(\frac{r-c_j}{w_j}\right)\right] \;\;,
\end{equation}

\noindent with the parameters  given in the 
Table \ref{table1}. Fig.~\ref{fig2} shows the potential
 versus distance in dimensionless units. In this work, we use $\epsilon/\gamma = 0.02$.

This effective potential was derived from the 
Ornstein-Zernike and integral equations applied to
the oxygen-oxygen radial distribution function of the atomistic model 
ST4~\cite{head_gordon_stillinger_1993}. At
 short distances it shows two characteristc length scales: one at 
$r_1\approx 1.1$ and another at $r_2\approx 1.6$, as 
in the usual coarse grained potentials proposed to
model the anomalous liquids.  In
addition a third length 
scale at $r_3\approx 2.2$ is also present. Since the 
derivation of the potential was based in the oxygen radial
distribution function these length scales represent the 
oxygen-oxygen distances 
related to the relevant coordination shells in the liquid.
The bulk system exhibits waterlike anomalies, and the liquid-gas
and liquid-liquid critical points predicted for 
water~\cite{ney_evy_marcia_2009}.

In the confined system the particles of the fluid
interact with the wall by the
 Weeks-Chandler-Andersen (WCA)
potential,
\begin{equation}\label{eq_potential3}
 U(r) = \left\{ \begin{array}{ll}
 U_{{\rm {LJ}}}(r) - U_{{\rm{LJ}}}(r_{cw})  , \qquad r \le r_{cw} \\
0   , \qquad r  > r_{cw} \;,
\end{array} \right.
\end{equation}
\noindent where $U_{LJ}(r)$ is the standard 12-6 LJ 
potential~\cite{AllenTild}. The
cutoff distance is $r_{cw} = 2^{1/6}$. For smooth plates, 
the interaction occurs just in $z$ direction and the potential is
written as $U_{LJ}(z)$.

 \begin{table} [!htb]
  \caption{Parameters of the particle-particle potentials in 
units of {\AA} and kcal/mol.} 
\vspace{0.5cm}
  \begin{tabular}{cc|cc|cc|cc}
  \hline\hline
\ \ Parameters \ & \ \ values  \ & \ \ Parameters \ & \ \ values \ \ & \ \ Parameters \ & \ \ values  \ \ & \ \ Parameters \ & \ \ values       \ \ \tabularnewline \hline
\ \ $a$       \  & \ \ $9.065$ \ & \ \  $w_1$  \    & \ \ $0.253$    & \ \  $h_1$  \    & \ \ $0.5685$    & \ \ $c_1$      \ & \ \ $2.849$      \ \ \tabularnewline
\ \ $b$        \ & \ \ $4.044$ \ & \ \  $w_2$  \    & \ \ $1.767$    & \ \  $h_2$  \    & \ \ $3.626$     & \ \ $c_2$      \ & \ \ $1.514$      \ \ \tabularnewline
\ \ $\epsilon$ \ & \ \ $0.006$ \ & \ \  $w_3$  \    & \ \ $2.363$    & \ \  $h_3$  \    & \hspace{-2mm} $-0.451$    & \ \ $c_3$      \ & \ \ $4.569$      \ \ \tabularnewline
\ \ $\sigma_p$ \ & \ \ $4.218$ \ & \ \  $w_4$  \    & \ \ $0.614$    & \ \  $h_4$  \    & \ \ $0.230$     & \ \ $c_4$      \ & \ \ $5.518$      \ \ \tabularnewline\hline\hline
 \end{tabular}\label{table1}
\end{table}


The dynamic, thermodynamic and structural properties of 
the fluid was studied using molecular dynamics 
simulation in the $NVT$ ensemble. The Nose-Hoover
thermostat was used to fix the temperature, with a coupling
 parameter $Q = 2$. The interaction potential between 
particles, Eq.~(\ref{eq_potential}), has a cutoff radius $r_c = 3.5$.

The fluid was confined by two different kinds of parallel walls: smooth 
and structured. The 
plates are fixed and
are located each one at $z = 0$ and $z = d$. The smooth plates are modeled 
by force 
fields in $z$ direction and
have no structure. The interaction between smooth plates and the fluid 
was done 
using the WCA 
(Weeks-Chandler-Andersen) potential, like shown in 
Eq.~\ref{eq_potential3}, but 
considering just the $z$ component. The structured plates are constructed 
by placing 
spherical 
particles of effective diameter $\sigma$ in a square lattice of area 
$L^2$. In this case the
interaction also is given by the WCA potential of the Eq.~\ref{eq_potential3}.

In $z$ direction the space occupied for the fluid was limited by the 
confining plates.
Due the excluded volume between the fluid near to the plates, the distance $d$ 
between them need to be corrected
to an effective distance $d_e$, that can be approach by 
$d_e \approx d-\sigma$~\cite{Ku05}.
Consequently, the effective density will be $\rho_e = N/(d_eL^2)$. The 
symbol $_e$ will 
be omitted in order to simplify the discussion. 

Systems with plate separations  $d = 2.5$, $4.2$, $5.2$ and 
$8.0$ were
analyzed. Several densities and temperatures were simulated
 to obtain the full phase diagrams for each case. For systems with 
$d = 2.5$, $4.2$ and $5.2$ $N = 507$ particles were
employed, while for $d = 8.0$ $N = 546$ particles were used.
Two different initial configuration of the systems were simulated: solid and 
liquid states. Using
different initial configurations allow us to identify precisely the final state 
of the system, avoiding metastability.
The equilibrium state was reached after $4 \times 10^5$ steps, followed 
by $8 \times 10^5$ simulation run.
We used a time step $\delta t = 0.001$, in reduced units, and all the 
physical quantities were get with $50$ uncorrelated samples.
To check the stability of the systems, we verify the energy as function 
of time and the perpendicular pressure
and parallel pressure as function of density.

Since the fluid is confined in the $z$ direction, the thermodynamic 
averages was 
calculated in 
components parallel and perpendicular to the plates~\cite{Me99}. The 
systems have 
periodic boundary conditions in $x$ and $y$ directions and they are extensive 
just in area and not in the distance between the plates. In this way,
only the parallel pressure might scale with the experimental pressure and the 
quantities of interest are related to parallel direction.

The parallel pressure was calculated using the Virial expression for the 
$x$ and $y$ directions~\cite{Ku05},
\begin{eqnarray}
\label{pressao_lateral}
P_{\parallel} = \rho k_B T + \frac{1}{2V}\left \langle 	\mathcal 
V _{\parallel} \right \rangle,
\end{eqnarray}
\noindent where $\mathcal V _{\parallel}$ is given by
\begin{eqnarray}
\label{pressao_lateral-2}
\mathcal V _{\parallel} = -\sum_{i=1}\sum_{j>i}\frac{x_{ij}^2+
y_{ij}^2}{r_{ij}}\left( \frac{\partial U(r)}{\partial r}\right )_{r=r_{ij}}.
\end{eqnarray}
The lateral diffusion coefficient, $D_{\parallel}$, was calculated using 
the mean 
square displacement (MSD), related from
Einstein relation,

\begin{eqnarray}
\label{difusao_lateral}
D_{\parallel} = \lim_{\tau\to\infty} 
\frac{\langle\Delta r_{\parallel}(\tau)^2\rangle}{4 \tau},
\end{eqnarray}
\noindent where $r_{\parallel} = (x^2+y^2)^{1/2}$ is the parallel 
distance of the particles.

The structure of the system was studied considering the lateral radial 
distribution function, 
$g_{\parallel}(r_{\parallel})$, calculated in specific slabs between the 
plates. The 
definition of
the $g_{\parallel}(r_{\parallel})$ is usually given by

\begin{eqnarray}
\label{gr_lateral}
g_{\parallel}(r_{\parallel}) \equiv \frac{1}{\rho ^2V}
\sum_{i\neq j} \delta (r-r_{ij})\left [ \theta\left( \left|z_i-z_j\right| 
\right) - \theta\left(\left|z_i-z_j\right|-\delta z\right) \right].
\end{eqnarray}

The $\theta(x)$ is the Heaviside function and it restricts the sum of 
particle pairs 
in the same slab of 
thickness $\delta z = \sigma$. The $g_{\parallel}(r_{\parallel})$ is 
proportional to the 
probability of finding a particle at a distance 
$r_{\parallel}$ from a referent particle.

\section{\label{sec:results} Results}

\subsection*{Thermodynamic, dynamic and structural 
behavior Smooth plates}

\begin{figure}[!htb]
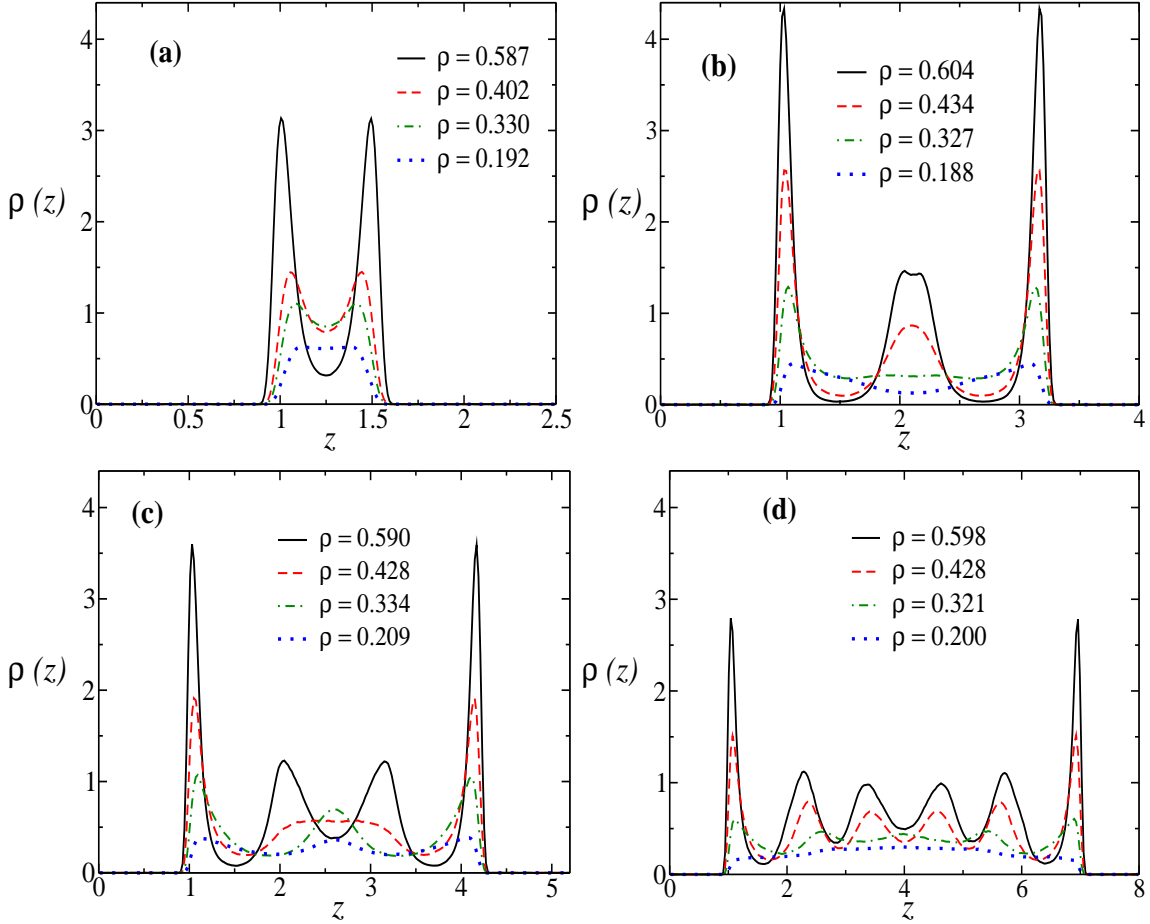

 \centering
 \begin{tabular}{cc}
 \includegraphics[clip=true,width=7.5cm, height=6.0cm]{fig3a.eps} 
 \includegraphics[clip=true,width=7.5cm, height=6.0cm]{fig3b.eps} \\
 \includegraphics[clip=true,width=7.5cm, height=6.0cm]{fig3c.eps}
 \includegraphics[clip=true,width=7.5cm, height=6.0cm]{fig3d.eps}
\tabularnewline
 \end{tabular}\par
 \caption{Transversal density profiles for systems confined by smooth plates 
with $T = 0.80$ and different densities at 
(a) $d = 2.5$, (b))$d = 4.2$ (c)$d = 5.2$ and (c) $d =8.0$. }
\label{fig3}
\end{figure}

A schematic depiction of the system confined by smooth plates is shown in 
the figure~\ref{fig1}(a). 
First, the effect of the structure of the liquid when confined by an 
uniform
field  is checked. Fig.~\ref{fig3} illustrates the transversal density 
profiles for plates separated at (a) $d = 2.5$,(b) $d = 4.2$ (c) $d = 5.2$ and (d) 
$d = 8.0$ at  $T = 0.80$ and several densities. 
In all cases the system form layers, however 
the number
of layers is dependent on the degree of confinement
and of the density of the systems, what is consistent with results
for atomistic such as SPC/E~\cite{Gi09a}
and coarse grained approximations with three body terms as mW~\cite{Mo12} 
model.

For higher degrees of confinement, $d = 2.5$, the fluid is structured in one 
or two layers,
depending on the density of the system. Fig.~\ref{fig3}(a) shows
 that two layers are 
observed for high densities ($\rho = 0.587$), while one layer occurs for 
low densities 
($\rho = 0.192$). The mechanism for 
the presence of different structures goes as follows.  For low densities
the wall does not induce correlations and layering at the $z$ direction
therefore one layer or bulk structure is formed.
As the density becomes higher, the competition 
between particle-particle and wall-particle interactions leads
to the formation of layers. Since the distance $d = 2.5$
does not allow formation of layers with distance $r_3-r_1\approx 1.1$
from each other (minimum energy), 
the layers are at a distant $r_2-r_1\approx 0.5$, which is the second
lowest energy potential.

For other degrees of confinement, $d = 4.2, 5.2, 8.0$, the same
competition between wall-particle and
particle-particle interactions appears
as shown in the Fig.~\ref{fig3}.  For 
low densities an uniform distribution with just
one layer appears and as the density
increases, two, three, four or even five layers are present. However,
since in this case the plates are further apart, the interlayer distance
is  is $r_3-r_1\approx 1.1$  that corresponds to the distance between
the shoulder length scale and the third coordination shell in the 
Fig. \ref{fig1}.  

\begin{figure}[!htb]
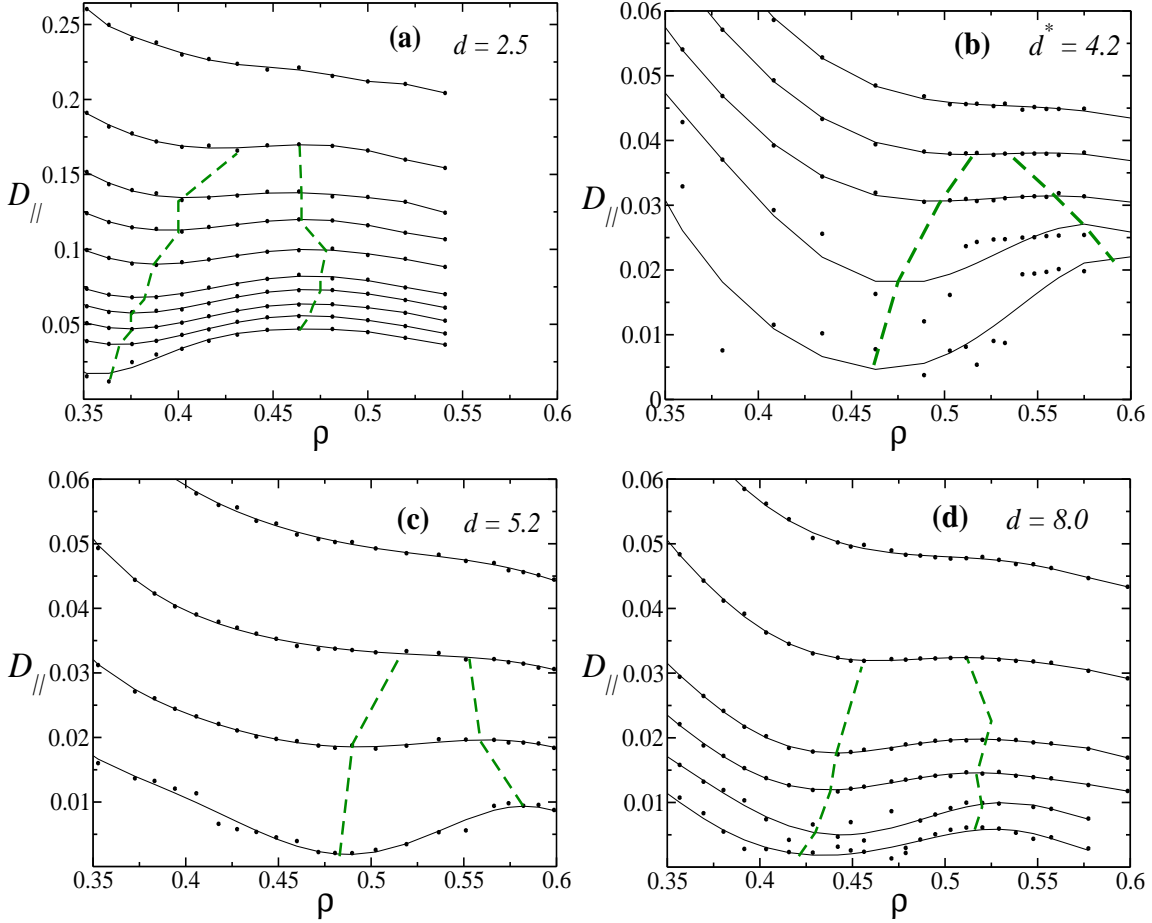

 \centering
 \begin{tabular}{cccc}
 \includegraphics[clip=true,width=7.5cm, height=6.0cm]{fig4a.eps}
\includegraphics[clip=true,width=7.5cm, height=6.0cm]{fig4b.eps}\\
 \includegraphics[clip=true,width=7.5cm, height=6.0cm]{fig4c.eps}
 \includegraphics[clip=true,width=7.5cm, height=6.0cm]{fig4d.eps}
\tabularnewline
 \end{tabular}\par
 \caption{Diffusion coefficient as function of density for (a) $d = 2.5$ 
and isotherms
$0.60$, $0.65$,..., $1.50$, (b) $d = 4.2$ and 
isotherms $0.60$, $0.65$,..., $1.50$ ,  (c) $d = 5.2$ and 
isotherms $0.50$, $0.55$,..., $0.80$ and 
(d) $d = 8.0$
and isotherms $0.45$, $0.50$,..., $0.80$. 
The dots are the simulated data and the black solid lines are 
polinomial fits. The dashed 
green lines 
bound the region where the diffusion are anomalous.}\label{fig4}
\end{figure}

The diffusion anomaly observed in liquid water is characterized by the 
increase of the diffusion
coefficient of the fluid when the pressure, or density, increases. For 
normal fluids, this
 coefficient decreases
when the fluid is compressed. The Fig.~\ref{fig4} shows the 
lateral diffusion 
coefficient
($D_{\parallel}$) as function of density of the system for (a) 
$d = 2.5$, (b) $d=4.2$, (c) $d = 5.2$ and 
(d) $d = 8.0$.
The range in temperature and density for which the anomaly in diffusion
is the same for the distances $d=4.2, 5.2, 8.0$ but
is different at $d=2.5$.  These two behaviors, one at $d=2.5$
and another at larger distances might be related with the 
different length scales involved in the close and larger 
distances as observed in the Figure~\ref{fig3}.

\begin{figure}[!htb]
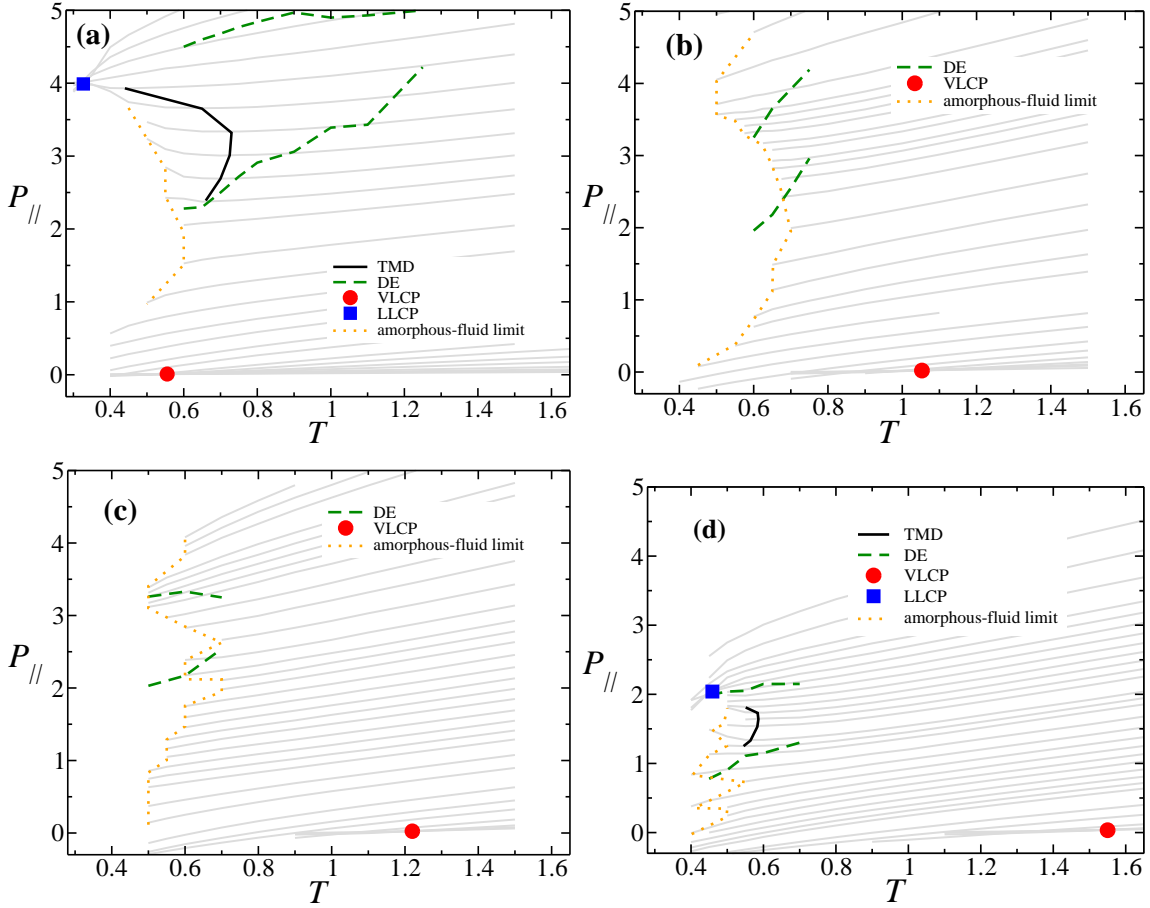

 \centering
 \begin{tabular}{cc}
 \includegraphics[clip=true,width=7.5cm]{fig5a.eps} 
 \includegraphics[clip=true,width=7.5cm]{fig5b.eps}\\
 \includegraphics[clip=true,width=7.5cm]{fig5c.eps}
 \includegraphics[clip=true,width=7.5cm]{fig5d.eps}
\tabularnewline
 \end{tabular}\par
 \caption{Parallel pressure versus temperature phase diagram for systems 
with smooth plates
separated by distances (a) $d = 2.5$, (b) $d = 4.2$,
(c) $d = 5.2$ and (d) $d = 8.0$ }\label{fig5}
\end{figure}

In addition to the anomalous dynamic properties of the 
confined liquid, the thermodynamic and phase space were also explored.
The system with $d = 2.5$ illustrated in 
the Fig.~\ref{fig5} (a) shows the 
presence of a Temperature of Maximum Density, $TMD$, as a solid line, a 
vapor phase and two liquid
phases. This system, therefore, exhibits  two stable critical points: a 
vapor-liquid
critical point, VLCP, at $P_c = 0.08$ and $T_c = 0.55$ (red circle) 
and a liquid-liquid 
critical 
point, $LLCP$, at $P_c = 4.0$ and $T_c = 0.3$ (blue square).
In the bulk system the  $VLCP$ occurs at $P_c = 0.078$ and $T_c = 1.98$ 
while the $LLCP$ appears at $P_c = 1.86$ and $T_c = 0.48$.
The comparison between the confined and the bulk systems indicates that 
the VLCP was shifted to lower temperatures, but did not present significantly 
changes in pressure. Meanwhile, the 
$LLCP$ is shifted to lower temperatures and higher pressures in relation to 
bulk, what is in agreement with results obtained for theoretical models involving 
anomalous fluids~\cite{truskett_2001}
and TIP4P water~\cite{Strekalova12b}. The dashed lines in the 
Fig.~\ref{fig5} (a) 
 represent the diffusion extremes and the 
pointed line indicates the limit between solid and fluid phases. The 
shifting of the critical point to 
lower temperatures can be assumed as a natural effect of the confinement, since the nanopore 
walls increase the entropy of 
the free energy of the system, favoring the disordered fluid phase. The 
increase in the 
pressure for the appearance of the $LLCP$ is the result of the layering 
imposed by the walls.
The layering allows for a high density interlayer making the full high 
density liquid only to appear at high densities.

For the plates separations $d = 4.2$ and $5.2$, the 
phase diagrams illustrated in the 
Fig.~\ref{fig5} (b) and (c)
show the presence of a 
VLCP also shifted to lower temperatures when compared
with the bulk system. However, the $TMD$ line and the $LLCP$
could not be determined. 

Due to the increase of 
the entropic effects for a system
under confinement 
 the melting line and the $LLCP$ should in 
principle move to 
lower temperatures. Whereas, we observed that the melting 
temperatures ($T_m$) for the confined systems 
are higher than the bulk system. In 
addition the change in the value of $T_m$
is not monotonic with $d$ similarly with 
what is observed in atomistic~\cite{De10}
and waterlike fluids~\cite{Kr13a}. 

\begin{figure}[!htb]
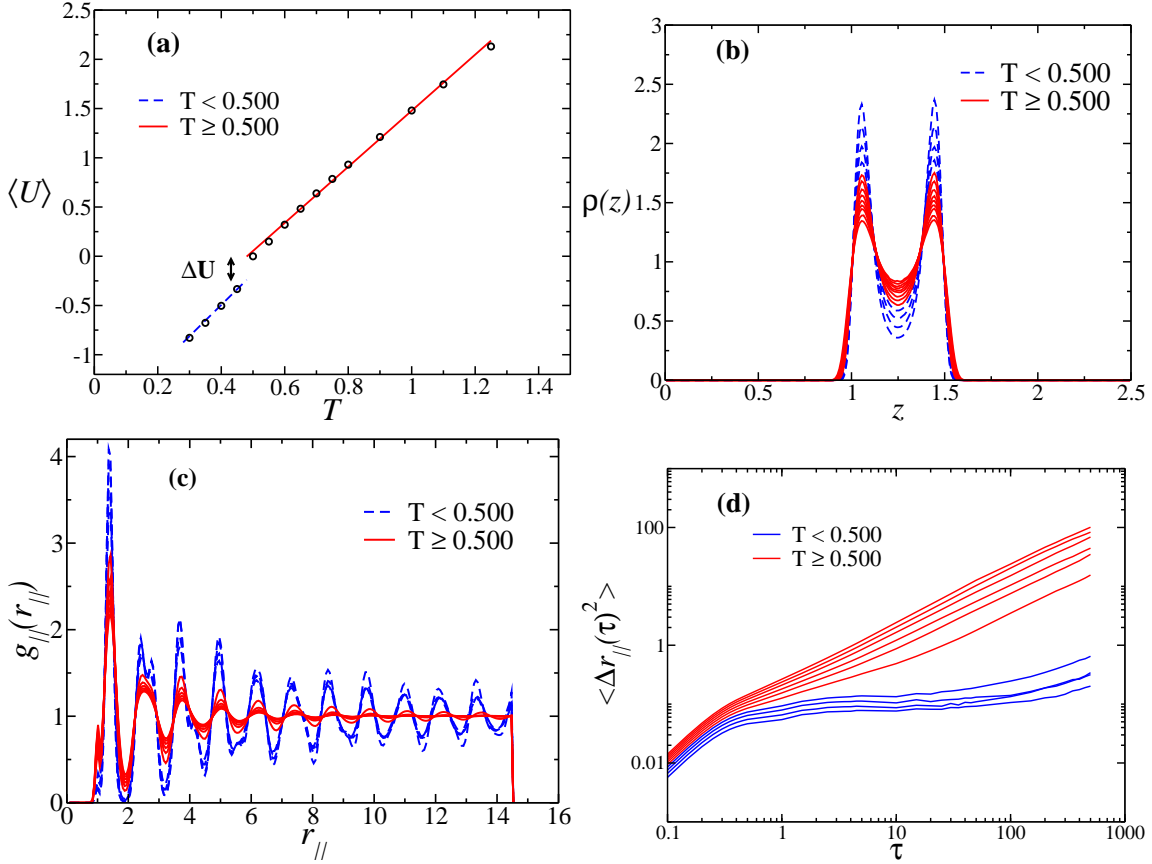

 \centering
 \begin{tabular}{ccc}
 \includegraphics[clip=true,width=7.5cm]{fig6a.eps} 
 \includegraphics[clip=true,width=7.5cm]{fig6b.eps}\\
 \includegraphics[clip=true,width=7.5cm]{fig6c.eps} 
 \includegraphics[clip=true,width=7.5cm]{fig6d.eps}
\tabularnewline
 \end{tabular}\par
 \caption{System with plates separated at $d = 2.5$ and density $\rho = 0.402$. In (a),
the mean potential energy as function of temperature, in (b) the transversal density profile,
in (c) the lateral radial distribution function ($g_{||}(r_{||})$) for the contact layer and
in (d) the mean square displacement in lateral direction.}\label{fig6}
\end{figure}

In order to understand why
the melting line moves to 
higher temperatures, covering
the $TMD$  and the $LLCP$, the structure
in this region was analyzed. For this
purpose, the transition
is analyzed for $d=2.5$ and $d=5.2$.
Fig.~\ref{fig6} in (a) illustrates  the 
mean potential energy as function of temperature,
in  (b) shows the transversal density profile, in 
(c) plots the lateral radial distribution funtion 
($g_{||}(r_{||})$) 
for the contact layer and in (d) presents the mean square 
displacement in lateral direction for  $d = 2.5$ and 
$\rho = 0.402$. We observe clearly a first order phase transition 
between a solid and a liquid phases. A discontinuous 
behavior was detected at $T = 0.50$. For $T < 0.50$, the energies 
have lower values, the density profiles and the 
$g_{||}(r_{||})$ have a well defined structure and the 
$\langle \Delta r^2(t)\rangle$ has a small inclination, 
showing a typical behavior of a solid/amorphous phase. Whereas, 
for $T\ge 0.500$, the energy shows high values
and the density profiles, the $g_{||}(r_{||})$ and the 
$\langle \Delta r^2(t)\rangle$ present a 
characteristic behavior of liquid phase. Solid-liquid first 
order phase transition was already observed
for TIP5P model confined between smooth hydrophobic plates~\cite{Han10}. 
The density profile shown in Figure.~\ref{fig6}(b) , however, 
indicates that solid phase is not structured inside each layer
but is present in the space between layers.

\begin{figure}[!htb]
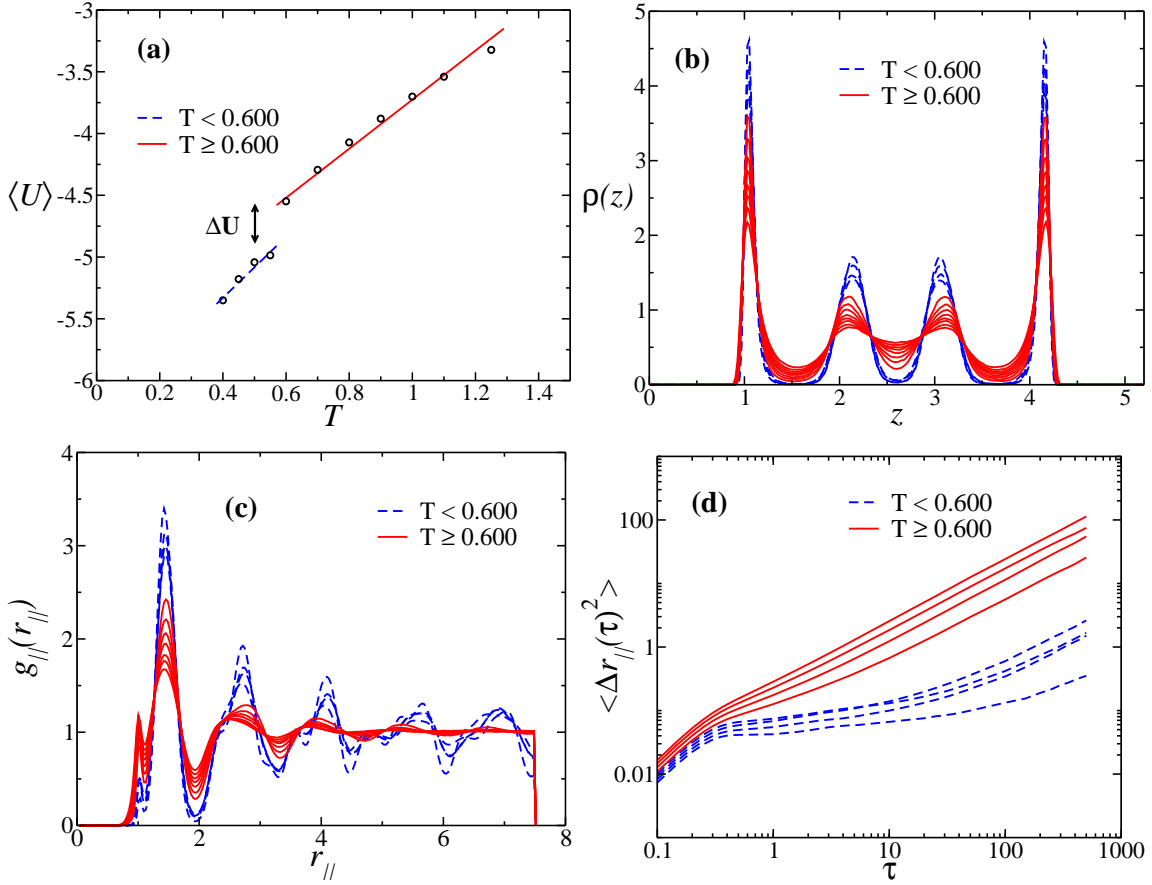

 \centering
 \begin{tabular}{ccc}
 \includegraphics[clip=true,width=7.5cm]{fig7a.eps} 
 \includegraphics[clip=true,width=7.5cm]{fig7b.eps}\\
 \includegraphics[clip=true,width=7.5cm]{fig7c.eps} 
 \includegraphics[clip=true,width=7.5cm]{fig7d.eps}
\tabularnewline
 \end{tabular}\par
 \caption{System with plates separated at $d = 5.2$ and density 
$\rho = 0.536$. In (a),
the mean potential energy as function of temperature, in (b) the 
transversal density profile,
in (c) the lateral radial distribution function ($g_{||}(r_{||})$) 
for the contact layer and
in (d) the mean square displacement in lateral direction.}
\label{fig7}
\end{figure}

In the case of $d=5.2$ illustrated in the Figures~\ref{fig7}
the first order transition is observed at $T=0.06$ because at
this temperature
the energy has a jump in (a),  the radial
distribution function shows a change in the structure from liquid
to solid in (c) and  the mean square displacement changes from
non zero to zero diffusion in (d). The density profile illustrated
in the  Figures~\ref{fig7} (b) differently than what
is observed for $d=2.5$ shows that the solid structure is 
confined to a single layer.

The different ways in which the solid structures accommodates for
the cases $d=2.5$ and $d=5.2$ under confinement explains
the non monotonic behavior of the melting temperature. While
for strong confinement the solid forms across the layers 
in the region  $d=5.2$ (and also  $d=4.2$) the solid
structure is confined to a single layer. As the distance increases
further the solid are again formed across layers approaching the
bulk structure.

\begin{figure}[!htb]
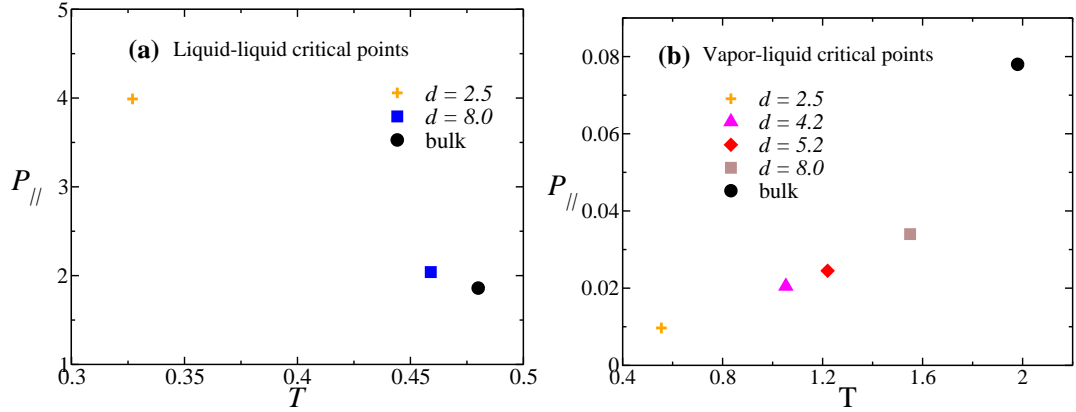

 \centering
 \begin{tabular}{cc}
 \includegraphics[clip=true,width=7cm]{fig8a.eps} 
 \includegraphics[clip=true,width=7cm]{fig8b.eps}
\tabularnewline
 \end{tabular}\par
 \caption{Location of (a) $LLCP$ and (b) $VLCP$ for all the distances between the smooth plates. }\label{fig8}
\end{figure}

The effects of the confinement in the critical points of water 
are dependent on the 
geometry and 
wall structure of confinement that are being considered. For exemple, when the 
water is confined in the porous matrix~\cite{Strekalova12b},
the $LLCP$ and the $TMD$ line are shifted to lower temperatures and higher 
pressures in relation
to bulk. But, in aqueous solutions 
of NaCl, Corradini and Gallo~\cite{CoG11} 
shows that the increase of salt concentration in water (TIP4P) shifts the 
$LLCP$ to higher temperatures and lower pressures
in relation to bulk. Our results for the $LLCP$ and $VLCP$ are summarized 
in Fig.~\ref{fig8} and are in good agreement with 
the results for the porous media~\cite{Strekalova12b}. This 
suggests that the salt/water long-range order
 interaction leads to changes in the water phase behavior what is 
  not present in the short-range wall-particle
interaction modeled by our system.

Experimental results show a non-monotonic behavior for the melting line 
and a strong dependence with the 
quality of the nanopore walls~\cite{De10}
what is observed in our results. In the next section we
will exam how 
the structure of the plates also have
important effect in the solidification of the system and in the location of 
the anomalies and critical points.

\subsection*{Structured plates}

\begin{figure}[!htb]
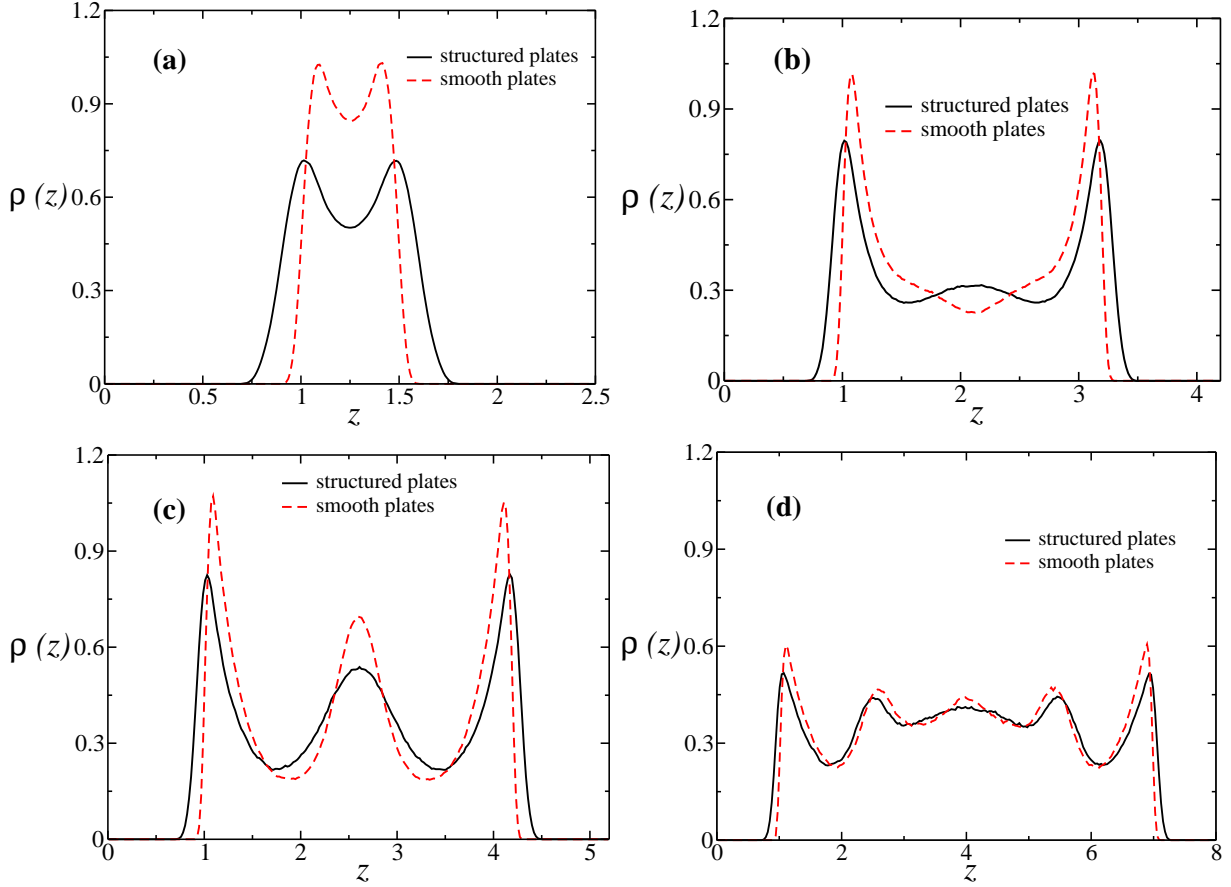

 \centering
 \begin{tabular}{cc}
 \includegraphics[clip=true,width=8cm]{fig9a.eps} 
 \includegraphics[clip=true,width=8cm]{fig9b.eps} \\
 \includegraphics[clip=true,width=8cm]{fig9c.eps}
 \includegraphics[clip=true,width=8cm]{fig9d.eps}
\tabularnewline
 \end{tabular}\par
\caption{Comparison of transversal density profile for systems confined by structured and
smooth plates at $T = 0.80$ and different densities at 
(a) $d = 2.5$, (b) $d = 5.2$ and (c) $d=8.0$. The confinement
at $d = 4.2$ is not shown for simplicity.}\label{fig9}
\end{figure}

The second scenario we address here
is the effect of the structure in the wall
has in the thermodynamic and dynamical
behavior of the confined liquid.  In this case, the 
plates are constructed by spherical particles in a square lattice,
as sketched in Fig.~\ref{fig1} (b). The interaction potential
between 
fluid particles and walls particles is given by the 
WCA potential (Eq.~\ref{eq_potential3}). A layering
structure similar to
picture observed for smooth plates analyzed in previous section is
also present for structured plates.
In Fig.~\ref{fig9} the transversal density
profiles for smooth and structured plates are compared for:
 (a) for $d = 2.5$ and $\rho = 0.310$,  
(b) for $d = 5.2$ and $\rho = 0.334$ and (c) for $d = 8.0$ and $\rho = 0.321$. In all these cases the temperature is the same,
$T = 0.80$. As the nanopore width decreases, the difference in the 
layer structure between the smooth and the structured walls 
increases. For $d = 8.0$, the fluid exhibits almost the same  
density profile for the two types  of confinement. This shows 
that for confined systems the fluid density profile
 is affected by the nanopore structure, particularly
for strongly confined systems.

\begin{figure}[!htb]
 \centering
 \begin{tabular}{ccc}
 \includegraphics[clip=true,width=8cm]{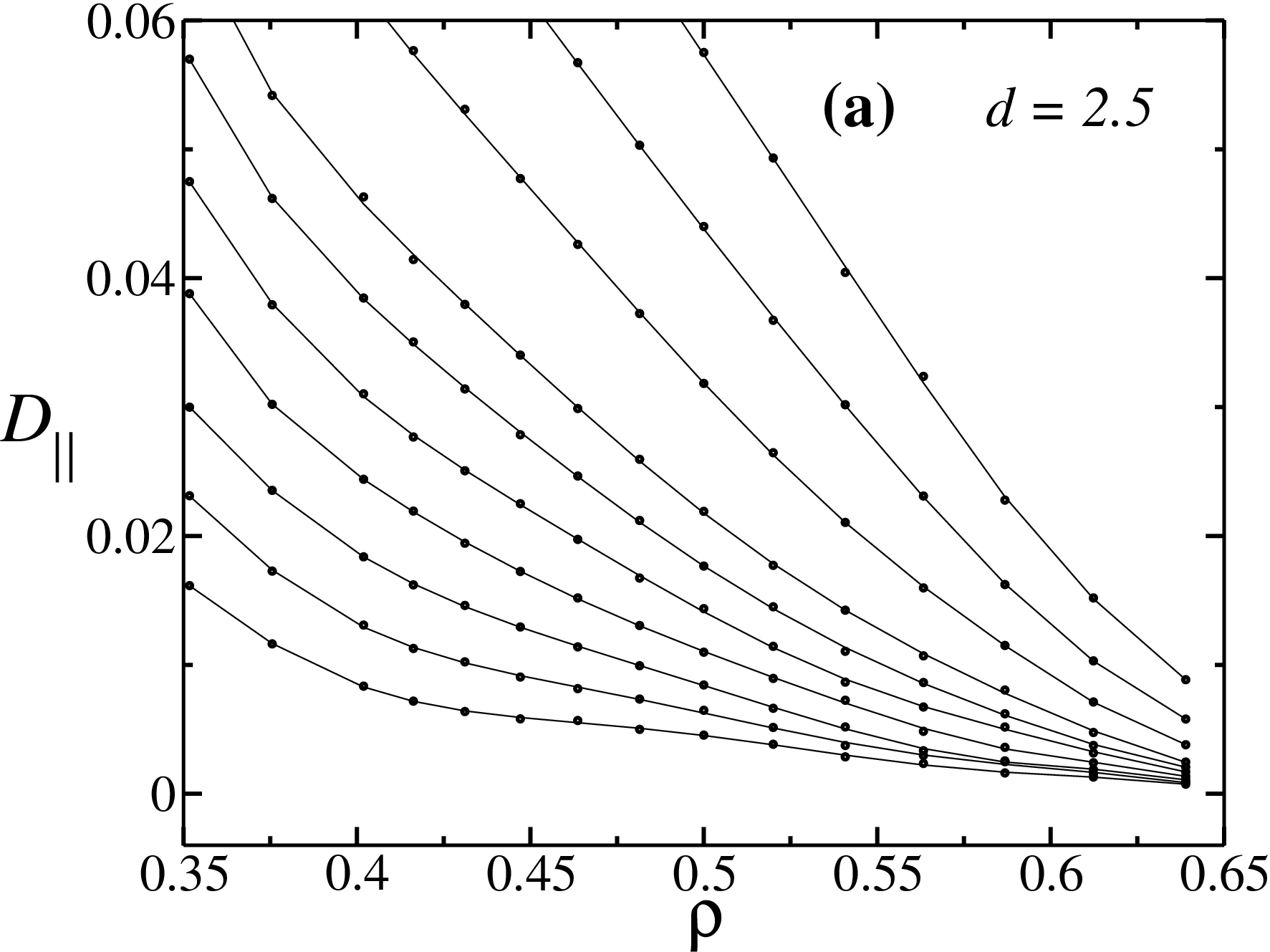} 
 \includegraphics[clip=true,width=8cm]{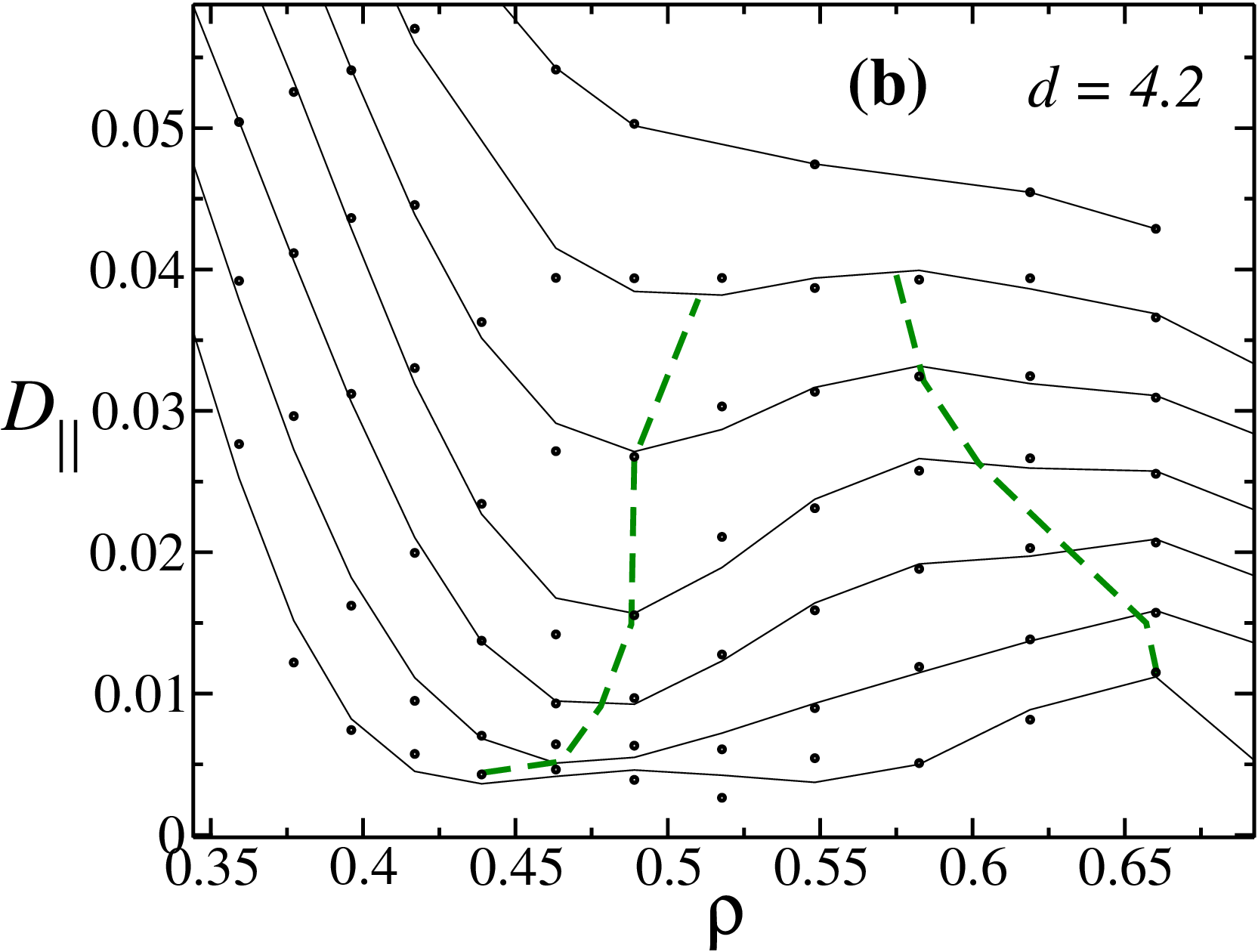} \\
 \includegraphics[clip=true,width=8cm]{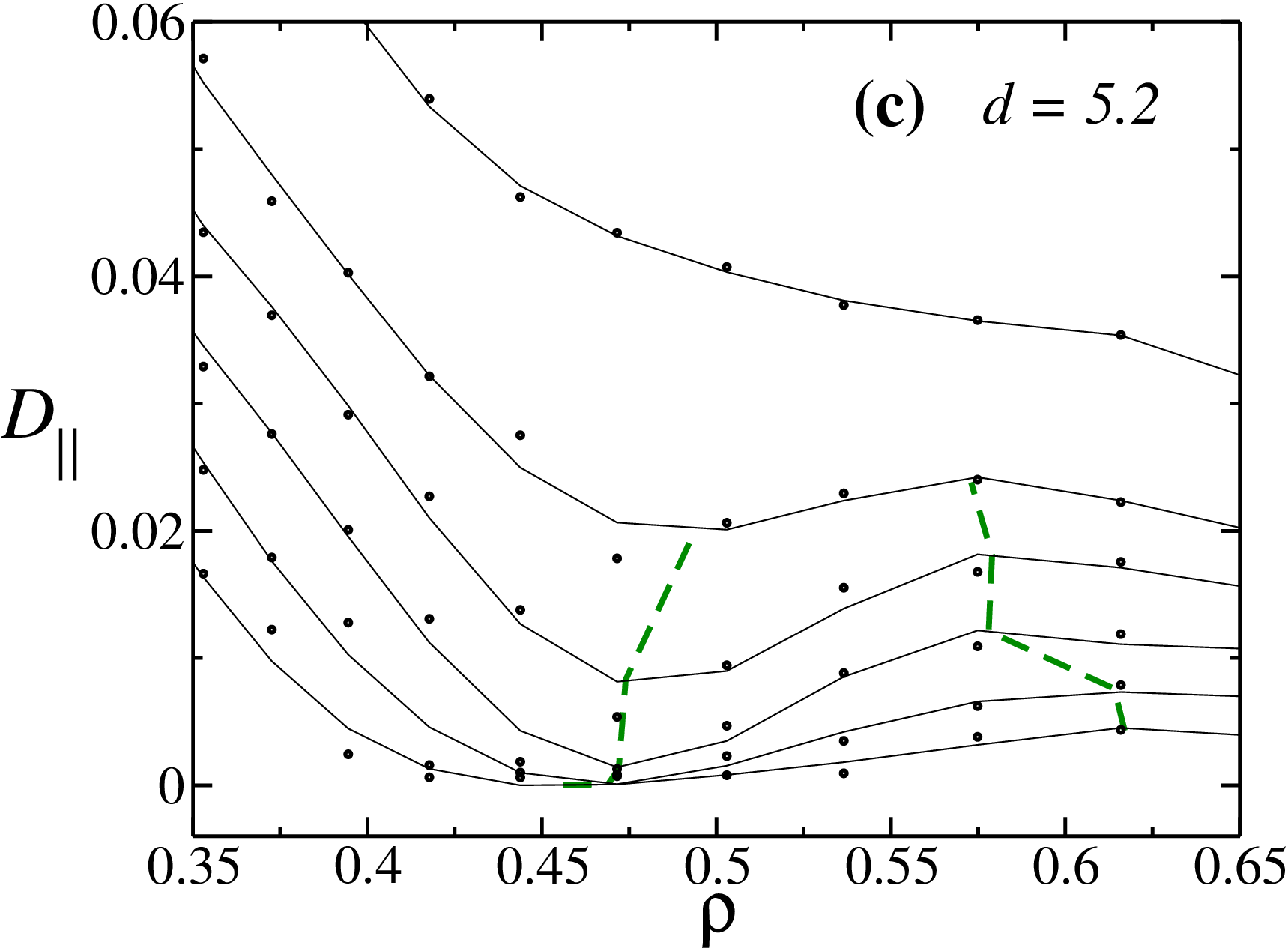}
 \includegraphics[clip=true,width=8cm]{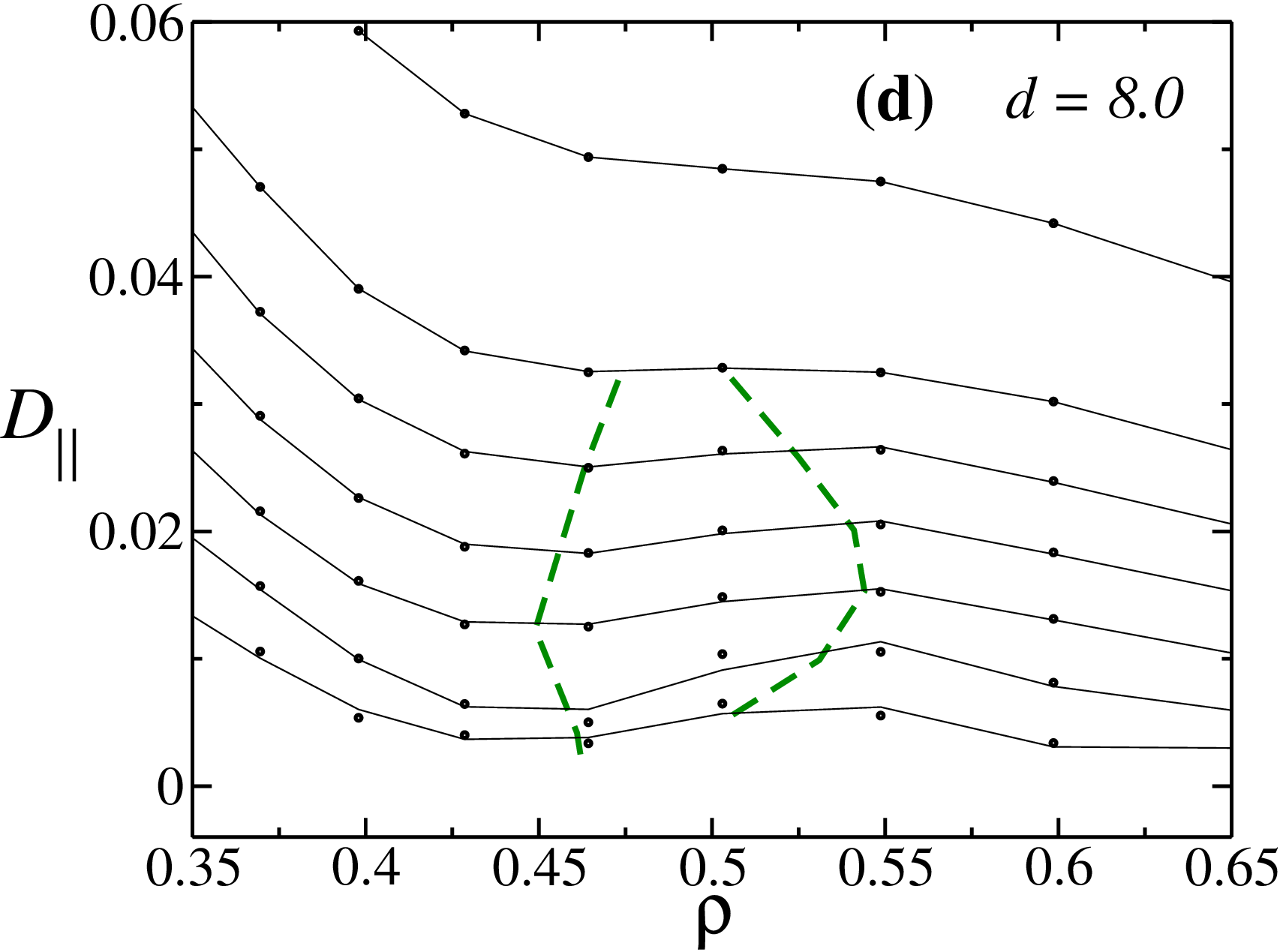}
\tabularnewline
 \end{tabular}\par
 \caption{Diffusion coefficient as function of density for (a) $d = 2.5$ and 
 isoterms $0.50, 0.55,..., 1.10$, (b) $d = 5.2$ and isoterms $0.40, 0.45,..., 0.80$ and (c) $d = 8.0$
 and isoterms $0.30, 0.40,..., 0.80$. The dots are the simulated data and the black solid lines are 
 polinomial fits. The dashed green lines bound the region where the diffusion are anomalous.}\label{fig10}
\end{figure}

Another property of the liquid
in which the structure of 
the confining surface might matter is 
the diffusion. Fig.~\ref{fig10} illustrates
the diffusion
coefficient in the parallel direction to the plates as function of 
the fluid density,
for nanopores with size $d = 2.5$, $d = 4.2$, $d = 5.2$ 
and $d = 8.0$. The 
diffusion anomaly was observed for 
systems with plates separated at $d = 4.2$,
$d = 5.2$ and  $d = 8.0$, while for $d = 2.5$ no 
anomalous behavior was detected
in the range of temperatures studied -- at 
low temperatures the fluid presents solidification, lead 
mainly by the nanopore structure.

Comparing the dynamical behavior of the systems, we 
verify that the fluid confined between structured
plates behaves completely different from the smooth cases, particularly
 for small values of $d$.
For $d = 2.5$, systems confined by smooth plates shows
a large region of pressures and temperatures
in which the diffusion anomaly is present 
 (Fig.~\ref{fig4}), while for structured walls, the fluid
dynamically behaves like normal systems, without 
diffusion anomaly for the range of temperatures studied.
The reason for this difference is that the structure of the wall
plays a very important role in the structure of the liquid
close to the wall and since at $d = 2.5$ the liquid
is closer to the wall when compared with the smooth plates, 
the structure the wall determines the arrangement of 
the liquid. The liquid particles will be able to occupy the 
space between the wall particles.

\begin{figure}[!htb]
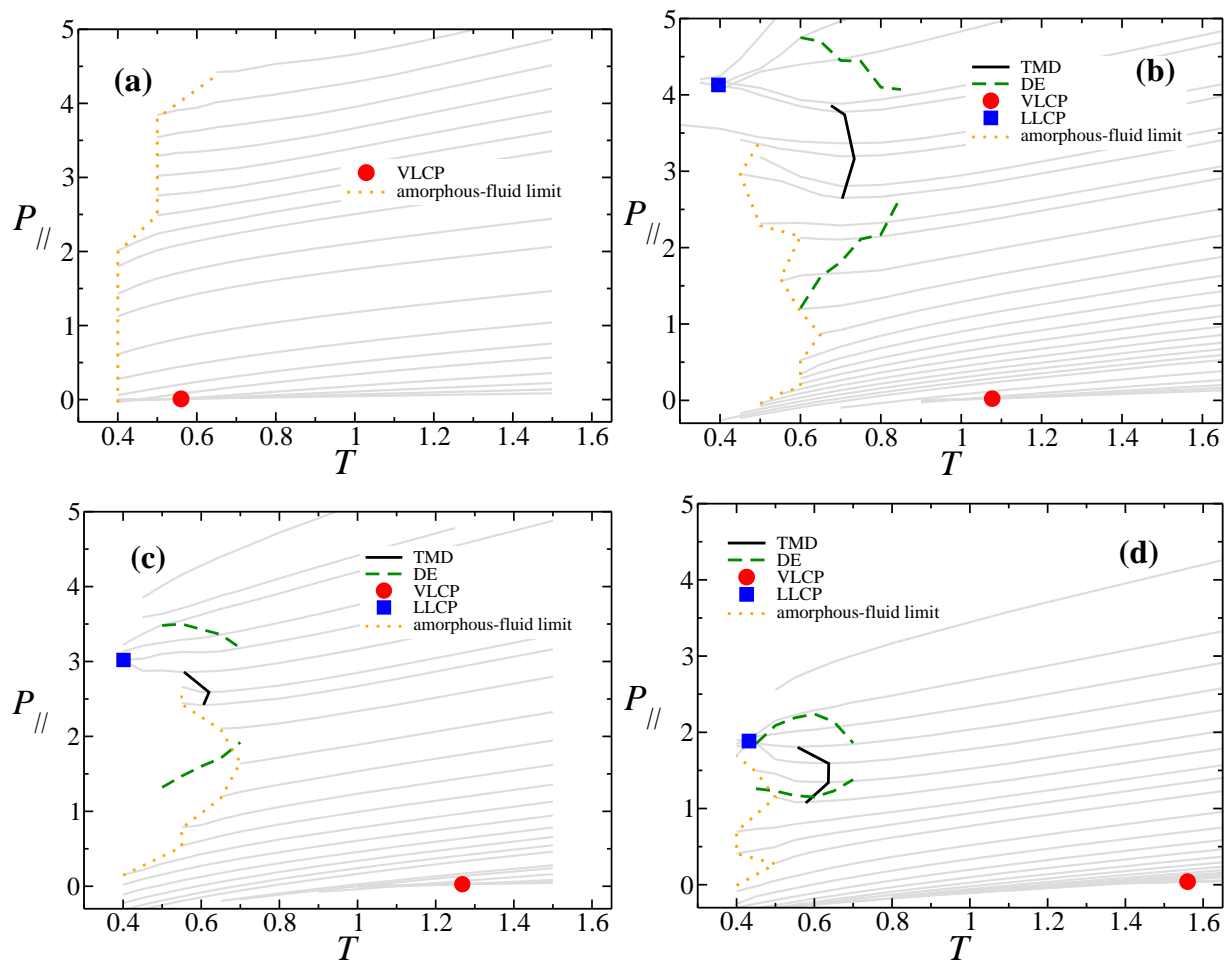

 \centering
 \begin{tabular}{ccc}
 \includegraphics[clip=true,width=8cm]{fig11a.eps} 
\includegraphics[clip=true,width=8cm]{fig11b.eps} \\
 \includegraphics[clip=true,width=8cm]{fig11c.eps}
 \includegraphics[clip=true,width=8cm]{fig11d.eps}
\tabularnewline
 \end{tabular}\par
 \caption{Parallel pressure versus temperature phase diagram for systems with structured plates
separated by distances (a) $d = 2.5$, (b) $d = 5.2$ and (c) $d = 8.0$. System
at $d = 4.2$ is not shown for simplicity. }\label{fig11}
\end{figure}

The  parallel  pressure versus 
temperature phase diagrams are shown in the Fig.~\ref{fig11}
for (a) $d = 2.5$, (b) $d = 5.2$, (c) $d = 4.2$ and (d) $d = 8.0$.
The lines in the graph
go as follows: the $TMD$ lines for each case is represented by
solid lines, the diffusion
extremes by dashed lines, the $VLCP$ by squares, the $LLCP$ by circles
 and the limit between fluid and solid phases by dotted lines. 

For structured nanopores with $d = 2.5$, the density and diffusion 
anomalies and the $LLCP$ 
are not observed outside the amorphous regions. This is an 
effect of the influence of the wall-water potential
that favors particles close to the wall to occupy the 
spaces between wall particles. Then the particle-particle two
length scales competition that leads to the presence
of density and diffusion anomalies does not happen, instead
there is a competition between particle-particle and wall-particle 
interactions. 
The solidification for the system in this case is similar to what happens 
in the last section for
$d = 4.2$, $d = 5.2$ and $d = 8$. In this
case, however, the  melting temperatures are
lower than in the smooth potential case.
The competition between the wall-particle
interaction that favors one solid arrangement
with the particle-particle interaction that
favor other arrangement explains the 
difference between the melting for 
rough and smooth walls. Classical water model TIP5P confined
between structured hydrophobic plates also
presents a shift for higher temperatures~\cite{zangi_2004}.

\begin{figure}[!htb]
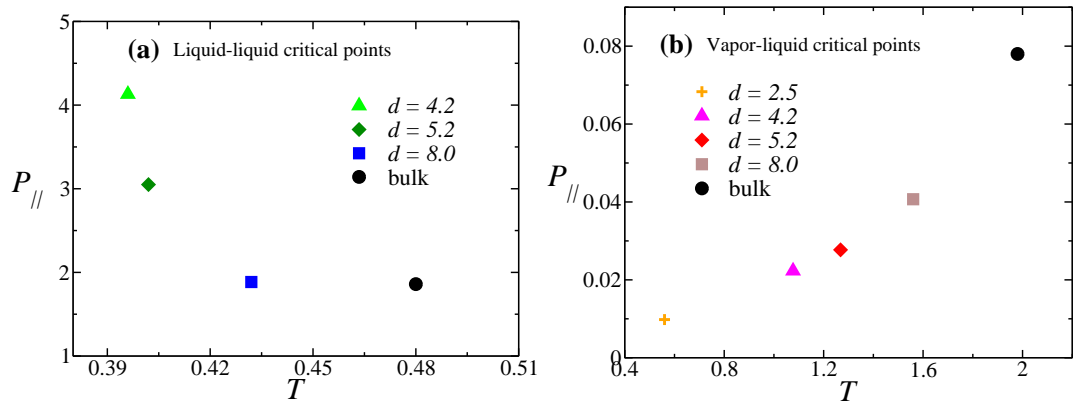

 \centering
 \begin{tabular}{cc}
 \includegraphics[clip=true,width=7cm]{fig12a.eps} 
 \includegraphics[clip=true,width=7cm]{fig12b.eps}
\tabularnewline
 \end{tabular}\par
 \caption{Location of (a) $LLCP$ and (b) $VLCP$ for all the distances between the structured plates. }\label{fig12}
\end{figure}

The Fig.~\ref{fig12} summarizes the behavior of (a) $LLCP$ and (b) $VLCP$ for the different
nanopores sizes and structured walls. The location of both critical points changes with the distances between the plates. 
As the nanopore width $d$ decreases, the $LLCP$ goes to lower temperatures and higher 
pressures, while the $VLCP$ is shifted to lower temperatures and lower pressures too.

\section{\label{sec:conclusions} Conclusions}

In this work we have studied the effects of the nanopore structure
and of the water potential length scales in the waterlike 
properties of a anomalous fluid. First, we 
tested the effect of using a three length scales
potential for analyzing the fluid behavior. In this case the system
confined by very small distances exhibits 
a different behavior when compared with 
confinement by intermediate and large distances.
This difference can be explained by 
the the arrangement of the fluid particles
in the first, second or third 
length scale of the potential. 
Then we check the differences in
the thermodynamic and dynamic anomalies of the  fluid 
when it was confined between smooth  
and structured walls. 
When observed, the density and diffusion anomalies are shifted to 
lower temperatures and
higher pressures in relation to bulk for both kinds of 
confinement. However, the critical
points and the limit between solid and fluid phases present a 
significant difference for
each system. For high degrees of confinement the properties of the fluid is
very well defined when confined by smooth nanopores, but the fluid 
crystallizes for structured walls
and small $d$. For intermediates separation of walls, smooth 
confinement present solidification
and structured confinement do not. So, a non-monotonic behavior is 
observed in the properties
of the fluid with $d$ when confined by smooth plates and a monotonic 
behavior with $d$
when confined between structured plates. The scales of the fluid-fluid and
the fluid-plate interaction potential are responsible for the different 
behavior observed
for each kind of confinement.

%
\section*{ACKNOWLEDGMENTS}
%
%
We thank for financial support the Brazilian science agencies, CNPq 
and Capes. This work is partially supported by CNPq, INCT-FCx.

\vspace{1cm}

\bibliographystyle{aip}

\end{document}